\documentclass[manuscript]{aastex}
\usepackage{natbib}
\usepackage{float}
\usepackage[caption = false]{subfig}

\shorttitle{PUIs in the Heliosphere and Heliosheath}
\shortauthors{Wu et al.}

\begin{document}

\title{Interstellar Pickup Ion Production in the Global Heliosphere and Heliosheath}

\author{Yihong Wu\altaffilmark{1,2}, Vladimir Florinski\altaffilmark{1,2}, and Xiaocheng Guo\altaffilmark{1}}

\altaffiltext{1}{Center for Space Plasma and Aeronomic Research, University of Alabama in Huntsville, Huntsville, AL 35805, USA; yw0009@uah.edu}
\altaffiltext{2}{Department of Space Science, University of Alabama in Huntsville, Huntsville, AL 35899, USA}

\begin{abstract}
Interstellar Pickup ions (PUIs) play a significant part in mediating the solar wind (SW) interaction with the interstellar medium. In this paper, we examine the details of spatial variation of the PUI velocity distribution function (VDF) in the SW by solving the PUI transport equation. We assume the PUI distribution is isotropic resulting from strong pitch-angle scattering by wave-particle interaction. A three-dimensional model combining the MHD treatment of the background SW and neutrals with a kinetic treatment of PUIs throughout the heliosphere and the surrounding local interstellar medium (LISM) has been developed. The model generates PUI power law tails via second-order Fermi process. We analyze how PUIs transform across the heliospheric termination shock (TS) and obtain the PUI phase space distribution in the inner heliosheath including continuing velocity diffusion. Our simulated PUI spectra are compared with observations made by {\it New Horizons}, {\it Ulysses}, {\it Voyager 1, 2} and {\it Cassini}, and a satisfactory agreement is demonstrated. Some specific features in the observations, for example, a cutoff of PUI VDF at $v = V_{SW}$ and a $f \propto v^{-5}$ tail in the reference frame of the SW, are well represented by the model.
\end{abstract}


\section{INTRODUCTION}

A neutral atom from the local interstellar medium (LISM) can be ionized while drifting into the heliosphere and turn into an interstellar pickup ion (PUI). The neutral atom can become ionized through charge exchange with an ion, photoionization by sun light or impact ionization by Solar Wind (SW) electrons. Charge exchange is the dominant mechanism of the three. In the SW frame, a newly created ion is ``picked up'' and starts to gyrate about the direction of the magnetic field. Subsequently they may be scattered into an isotropic distribution by either ambient preexisting or self-excited waves \citep{1976JGR....81.1247V,1987JGR....92.1067I,1999SSRv...89..413Z}. The resulting PUI velocity distribution function (VDF) in the reference frame of the SW is a spherical shell, at the lowest order. These PUIs are also convected outward by the expanding SW. As the PUIs travel outward through the heliosphere, the shell becomes filled by adiabatic cooling. Generally, one can easily identify PUIs by their VDFs which are distinctly different from that of SW ions.

In view of the paucity of measurements of keV-ions in the outer heliosphere, the transport of PUIs is not yet fully understood. Especially, the production of the suprathermal (high-energy non-Maxwellian) tails on the VDFs is a subject of much debate \citep{2003JGRA..108.1266C,2011A&A...533A..92F}. It was suggested that during their propagation through the outer heliosphere, PUIs experience pitch-angle scattering and stochastic acceleration by interactions with different kinds of SW turbulences \citep{1997A&A...320..659C}. 

PUI kinetic transport theory describes the evolving PUI distribution in phase space. The transport equation, an equation of ``motion'' for the distribution function, is the basis for almost all work on PUI transport. An early paper by \cite{1976JGR....81.1247V} calculated an isotropic PUI VDF without energy diffusion but including the effects of adiabatic deceleration. The VDF appears as a thick shell centered around the SW beam. \cite{1987JGR....92.1067I} investigated how PUI distribution varies with radial distance and phase space speed. He has shown that the effects of energy diffusion can be important and can lead to substantial particle acceleration. \cite{1995A&A...304..609C} studied the influence of different representations for the energy diffusion coefficient which would result from different radial variations of relative fluctuation amplitudes of MHD turbulence in the outer heliosphere. These authors have calculated the energy spectra of PUIs upstream of the heliospheric termination shock (TS) on the basis of realistic PUI production rates. They showed that second-order Fermi acceleration by means of Alfvenic turbulence produces the suprathermal tail. \cite{2011A&A...533A..92F} replaced the adiabatic cooling with the ``magnetic cooling'' process resulting from the conservation of the first and second adiabatic invariants. They have demonstrated that small second-order Fermi acceleration and magnetic cooling generates a PUI VDF with $f\propto v^{-5}$. \cite{2012JGRA..117.6104I} first compared simulated PUI densities and temperatures with the SWICS measurements on board {\it Ulysses}. The model results matched well with the observations at about $5.2$ AU during both quiet periods and the disturbed period during the Halloween 2003 storm. Assuming that the PUI distribution functions are $\kappa$ distributions, \cite{2014JGRA..119.7998F} proceeded from the phase space transport equation to a pressure equation for the $\kappa$ parameter $\kappa = \kappa(r)$ as function of heliocentric distance $r$. They obtained the range of possible radial variations of $\kappa$ from relatively high values in the inner heliosphere to values between 1.5 and about 2 further out, depending on diffusion coefficient.

The purpose of the present paper is to place PUI transport in the context of the global picture of the SW-LISM interaction. The treatment is based on the more conventional view where wave-particle interactions ensure rapid isotropization of the distribution. A parallel can be drawn with the approach of \citet{2012ApJ...757...74G}, who used a grid-based model for the isotropic PUI distribution function and for the waves generated by the PUI ring anisotropization process, but neglected PUI momentum diffusion. Their model was applied to the supersonic SW upstream of the TS. Here we investigate the detailed spatial distribution of PUIs in both the supersonic SW and the heliosheath. We take into account the effects of convection with the SW, adiabatic cooling, second-order Fermi process and ionization. Similar to \citet{2012ApJ...757...74G}, the model combines the MHD treatment of the background SW and neutral atoms with a kinetic treatment of PUIs in the isotropic approximation.

The rest of this paper proceeds as follows. Section 2 explains how we simulate the SW-LISM interactions. The PUI transport model is introduced in Section 3. Section 4 presents the simulated PUI phase space density and energy spectra. Finally, Section 5 mentions the weaknesses of the presented model and proposes several improvements to the current methods.

\section{THE SW-LISM INTERACTION}

The heliosphere is a low-density bubble embedded in the local interstellar cloud. The SW slows down when interacting with the LISM. At the TS, a standing shock wave, the SW speed falls below the effective speed of sound (that includes a contribution from PUIs) and becomes subsonic. The TS causes compression, heating, and a change in magnetic field. The SW continues to slow down as it passes through the heliosheath, a transitional zone bounded by the TS and the heliopause, where the pressures from the LISM and SW are balanced. The LISM pressure causes the heliosphere to develop into a comet-like structure. The nose of the heliopause is defined as the direction opposite to the sun's motion through the Local Interstellar Cloud.

We use a three-dimensional MHD model with four neutral hydrogen atom populations to obtain the plasma background. The neutral hydrogen atoms are separated into four species according to the region in which they were created. Studies comparing multiple neutral fluid and the kinetic Monte Carlo approaches to model neutrals have been done and the 4-fluid model shows excellent agreement for the plasma properties \citep{1995JGR...10021595P,1996JGR...10121639Z,2006JGRA..111.6110H}. The four neutral fluids have distinct properties. The Neutral 1 population is the primary interstellar atoms. The Neutral 2 population consists of atoms born in the outer heliosheath. Neutral gas produced in the inner heliosheath constitute the Neutral 3 population, while Neutral 4 is created in the supersonic SW. The computer model is based on a hexagonal spherical geodesic grid that ensures a uniform partitioning of the surface of a sphere, combined with a concentric nonuniform radial grid \citep{2013ApJS..205...19F}. We solve a modified set of MHD equations using a finite volume method on this grid. We use 40,962 Voronoi polygons on the sphere and $528$ radial shells. Because most PUIs are produced at small heliocentric distances and convected outward, we placed the inner boundary at $1.5$ AU and the outer boundary at $800$ AU. The radial cells are smaller near the origin; their width increasing monotonically with radial distance. The $z$-axis points northward of the solar equator, and the $x$-axis is in the plane defined by the interstellar helium flow direction \citep{2005Sci...307.1447L} and the $z$-axis. The $x, y, z$-axes constitute a right-handed orthogonal system.

The model heliosphere corresponds approximately to the solar minimum conditions. We assumed the slow wind latitudinal extent angle of 36$^\circ$. At $1$ AU, in the slow SW, the bulk velocity $u=430$ km s$^{-1}$ and density $n=5.0$ cm$^{-3}$; in the fast SW, $u=725$ km s$^{-1}$ and $n=1.4$ cm$^{-3}$ \citep{2009JGRA..114.1109E,2011SoPh..274..321J}. At $1$ AU, the radial magnetic field component $B_r=24.5$ $\mu$G. The azimuthal magnetic field component $B_\phi \propto 1/u$, where $u$ is the SW speed. Since the heliospheric current sheet is too thin to affect the dynamics of the plasma on large scales, we do not include it in our simulation and use a unipolar magnetic field model \citep{2010A&A...516A..17C,2011ApJ...728L..21B,2015ApJS..220...32I,2015ApJ...800L..28O}.

The interstellar neutral atom density was $0.11$ cm$^{-3}$ and its velocity vector was $(-25.9, 0, -2.196)$ km s$^{-1}$ in our coordinates \citep{2012Sci...336.1291M}. We assumed the interstellar magnetic field direction toward the center of the {\it Interstellar Boundary Explorer (IBEX)} ribbon, with Cartesian components $(1.834, -1.866, 1.422)$ $\mu$G \citep{2013ApJ...776...30F}. The interstellar plasma and neutral temperatures were both equal to 6300 K. The plasma interacts with the hydrogen atoms through charge exchange. These atoms are modeled using a gas-dynamic model in a fashion similar to our MHD system except that the magnetic field is set to be zero. We use the charge exchange terms in \cite{1995JGR...10021595P}. The MHD code is run until a steady state is reached, which provides the plasma-neutral simulation background for the next step.

\section{THE PUI TRANSPORT MODEL}
The PUIs are non-thermal, and their preferred treatment is kinetic. We assume that PUIs are scattered into isotropic distribution by either ambient preexisting or self-excited waves, so the VDF depends only on the absolute value of velocity (in the plasma frame).

\cite{1996GeoRL..23.2871S} obtained the mean field strength $B_0$ from 1-day averages measured by {\it Ulysses}, and averaged the square of the variations over 5 days, $\eta^2=\langle(B-B_0)^2/B_0^2\rangle$, during the 150 days from February 19 to July 18, 1992. They found that {\it Ulysses} crossed flux tubes with different values of $\eta^2$, in the range from a high value of $\eta_{H}^2 \approx 0.12$ (highly turbulent) to a low value of $\eta_{L}^2 \approx 0.005$ (almost laminar). As long as the observed distribution functions are averaged over more than about 10 days, they can be viewed as a superposition of these high and low values of $\eta^2$. Following this idea, we separate the PUI distribution into two components. The first component (``core") is found in laminar flux tubes, while the second component (``tail") resides in turbulent flux tubes, where they experience stochastic acceleration.

The distributions $f_\mathrm{core}$ and $f_\mathrm{tail}$ satisfy the following transport equations:
\begin{equation}
\frac{\partial f_\mathrm{core}}{\partial t} + {\bf u} \cdot \nabla f_\mathrm{core}-\frac{\bf{\nabla\cdot  u}}{3} \frac{\partial f_\mathrm{core}}{\partial \ln v} = \Gamma * S,
\end{equation}
\begin{equation}
\frac{\partial f_\mathrm{tail}}{ \partial t} + {\bf u} \cdot \nabla f_\mathrm{tail}- \frac{\bf{\nabla\cdot u}}{3} \frac{\partial f_\mathrm{tail}}{\partial \ln v} -\frac{1}{v^2}\frac{\partial}{\partial v } \left( v^2 D \frac{\partial f_\mathrm{tail}}{\partial v}\right) = (1-\Gamma) * S
\end{equation}
with $\Gamma=0.98$, where $v$ is PUI velocity in the non-inertial frame moving with the plasma, $\bf u$ is the bulk velocity of the SW plasma, $D$ is the velocity diffusion coefficient, and $S$ is the source term, describing PUIs produced from charge exchange or some other ionization process. The left-hand sides of the transport equations describes the explicit time dependence of the PUI velocity distribution, convection, adiabatic cooling or heating and velocity diffusion, respectively. PUIs typically have energies between $100$ eV and $100$ keV. We neglect spatial diffusion and drift motions since the typical spatial diffusive scale of PUI is much smaller than our simulation scale, and the drift speed is small in this low energy range \citep{1993P&SS...41..773R}.

The velocity diffusion coefficient is $D=v^2\delta u/(9\xi_c)$, where $\delta u$ is the RMS fluctuating velocity of the plasma and $\xi_c$ is the correlation length \citep{1997A&A...320..659C}. It includes a multitude of effects, such as transit-time damping \citep{1976JGR....81.4633F}, parallel electric fields in two-dimensional turbulence \citep{2002JGRA..107.1138L}, and large-scale compressive structures \citep{2003JGRA..108.1266C}. The values of $\delta u$ and $\xi_c$ were ($29$ km s$^{-1}$, $0.3$ AU) in the inner heliosheath and ($24$ km s$^{-1}$, $1$ AU) in the supersonic SW. These parameters were chosen to produce suprathermal tails in a way that would be consistent with the {\it Voyager} observations (see below).

In the source term, charge exchange and photoionization are taken into account, but electron impact ionization is neglected. The distribution of PUIs (variables bearing a subscript ``i'') created by the charge exchange process holds imprints from the parent atom distribution (index ``n'') and the background plasma (index ``p''). We assume that both are Maxwellian with densities $n$, mean velocities ${\bf u}$ and thermal speeds $v_T=(2kT/m)^{1/2}$, 
\begin{equation}
f_p({\bf v}_p)=\frac{n_p}{\pi^{3/2}v^3_{T_p}}e^{-({\bf v}_p-{\bf u}_p)^2/v^2_{T_p}},
\end{equation}
\begin{equation}
f_n({\bf v}_n)=\frac{n_n}{\pi^{3/2}v^3_{T_n}}e^{-({\bf v}_n-{\bf u}_n)^2/v^2_{T_n}}.
\end{equation}
The PUI production term from charge exchange is
\begin{eqnarray}
\left(\frac{\delta f_i(v_i)}{\delta t}\right)_{ex} = \int \sigma(|{\bf v}_p-{\bf v}_n|)f_p({\bf v}_p)f_n({\bf v}_n)|{\bf v}_p-{\bf v}_n|\frac{d^3 v_p d^3 v_n}{d^3 v_i}
\nonumber \\
=\sigma(\Delta v_p({\bf v}_i))f_n({\bf v}_i)\int f_p({\bf v}_p)|{\bf v}_p-{\bf v}_i|d^3 v_p
\\
= n_p\sigma(\Delta v_p({\bf v}_i))\Delta v_p({\bf v}_i)f_n({\bf v}_i),
\nonumber
\end{eqnarray}
where $\Delta v_p({\bf v}_i)$ is the average relative speed between a given PUI and the proton population. To obtain the above equation, the condition for charge exchange, ${\bf v}_n={\bf v}_i$, and the experimental result that the cross section $\sigma(v)$ is only weakly dependent on the relative velocity between the particles, were used. Numerical values of $\sigma$ are available from \cite{2005JGRA..11012213L}. The average relative speed is computed as \citep{Fahr1967}
\begin{eqnarray}
\Delta v_p({\bf v}_i)=\frac{1}{\pi^{3/2}v^3_{T_p}}\int |{\bf v}_p-{\bf v}_i| e^{-({\bf v}_p-{\bf u}_p)^2 / v^2_{T_p}} d^3 v_p
\nonumber \\
= v_{T_p}\left[\frac{1}{\sqrt{\pi}} e^{-x^2}+\left(x+\frac{1}{2x}\right) \textnormal{Erf}(x)\right],
\end{eqnarray}
where $x=|{\bf v}_i-{\bf u}_p| / {v_{T_p}}$.

The PUI transport equation is solved in the plasma frame, so we transform (5) into that frame and average over a sphere. We introduce the PUI velocity in the non-inertial frame moving with the plasma, ${\bf v'}_i={\bf v}_i-{\bf u}_p={\bf x}v_{T_p}$. Notice that in this frame (6) is isotropic and requires no averaging. Then
\begin{eqnarray}
\left(\frac{\delta f_i(v'_i)}{\delta t}\right)_{ex} = \frac{n_p n_n \sigma(\Delta v_p(v'_i)) \Delta v_p(v'_i)}{4\pi^{5/2}v^3_{T_n}} \int e^{-({\bf v'}_i-{\bf u}_n+{\bf u}_p)^2/{v^2_{T_n}}}d\Omega'_i
\nonumber \\
=\frac{n_p n_n \sigma(\Delta v_p(x)) \Delta v_p(x)}{2\pi^{3/2}v^3_{T_n}} \int\limits_{-1}^{1} e^{-(\alpha^2x^2-2\mu'\alpha x h+h^2)}d\mu'
\nonumber \\
=\frac{n_p n_n \sigma(\Delta v_p(x)) \Delta v_p(x)}{4\pi^{3/2}v_{T_p}v^2_{T_n}} \frac{e^{-(\alpha x - h)^2}-e^{-(\alpha x + h)^2}}{xh},
\end{eqnarray}
where $x=v'/v_{T_p}$, $h=|{\bf u}_n-{\bf u}_p|/v_{T_n}$ and $\alpha=v_{T_p} / v_{T_n}$, is the charge exchange source term for an isotropic distribution of PUIs. Similarly, the photoionization source term is
\begin{eqnarray}
\left(\frac{\delta f_i(v'_i)}{\delta t}\right)_{ph}=\frac{n_n\nu}{4\pi^{3/2}v_{T_p}v^2_{T_n}} \frac{e^{-(\alpha x - h)^2}-e^{-(\alpha x + h)^2}}{xh},
\end{eqnarray}
where $\nu=\nu_{ph}(1\;\mathrm{AU}/r)^2$ with $\nu_{ph}$ being the rate of photoionization at $1$ AU.

We solve the PUI transport equations (1), (2) by integrating it on a multi-CPU cluster simultaneously with the MHD system. A conservative form of these equations is used, namely
\begin{eqnarray}
 \frac{\partial f}{ \partial t} + {\bf\nabla} \cdot({\bf u}f) - \frac{1}{v^2}\frac{\partial}{\partial v} \left[\frac{v(\bf{\nabla \cdot  u})}{3}v^2f + Dv^2 \frac{\partial f}{\partial v}\right] = S,
\end{eqnarray}
and similar for equation (2). The transport module is implemented on the same grid as the MHD simulation, using a finite volume method. Right and left interface values are computed using piecewise linear reconstruction with a WENO limiter \citep{1996jcoph.126..202J,1998jcoph.144..194F}. The ionic components all have the same bulk speed available from the MHD solution. The velocity grid extends from 10 km s$^{-1}$ to 6000 km s$^{-1}$.

\section{SIMULATION RESULTS}

\begin{figure}
\epsscale{1.0}
\plottwo{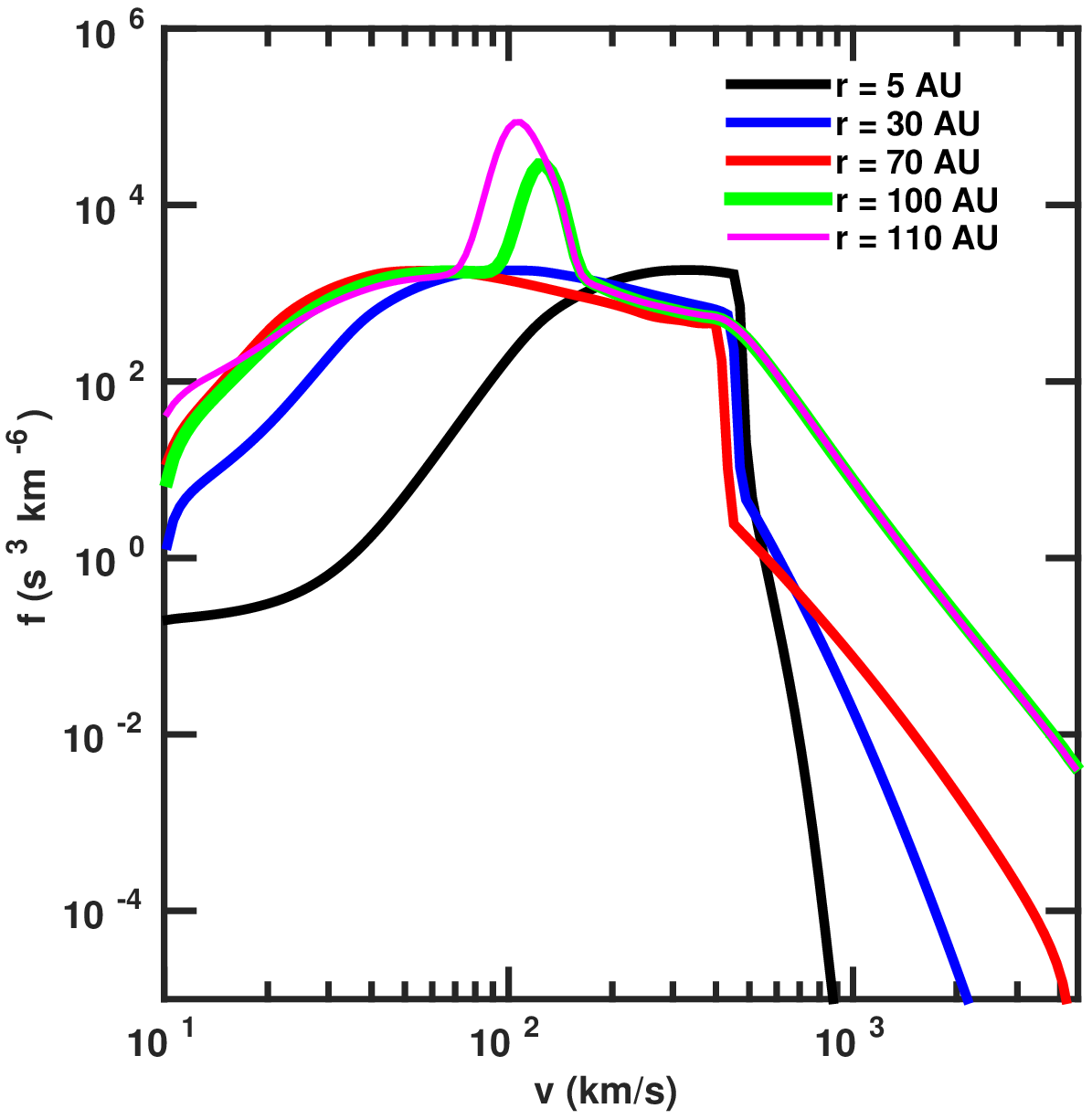}{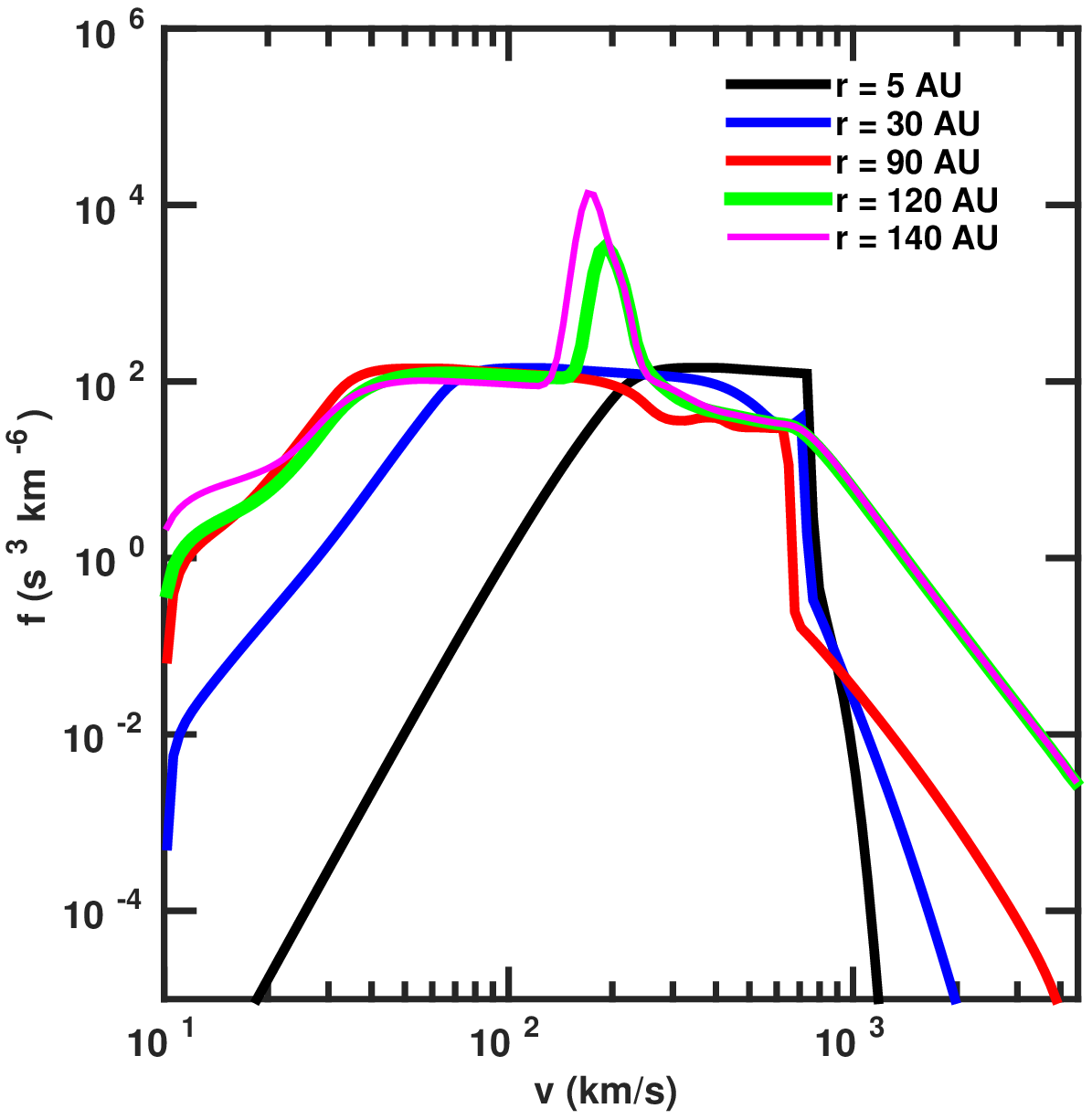}
\caption{PUI velocity distribution functions in the plasma frame along the {\it Voyager 1} direction ($\theta=55^{\circ}, \phi=0^{\circ}$) (left panel) and along the north polar direction ($\theta=0^{\circ}$) (right panel).}
\label{fig:1}
\end{figure}

The TS and the heliopause are at $90$ AU and $144$ AU respectively along the {\it Voyager 1} spacecraft direction in our simulations. {\it Voyager 1} actually crossed the TS and the heliopause at $94$ AU and $122$ AU \citep{2005Sci...309.2017S,2013Sci...341..150S}. These distances are appropriate for a model that is time independent and is based on solar-minimum conditions. In Figure \ref{fig:1}, the left panel shows the simulated PUI VDF in the {\it Voyager 1} direction ($\theta=55^{\circ}, \phi=0^{\circ}$, where $\theta$ denotes the co-latitude and $\phi$ denotes the longitude), in the plasma frame. At $r = 5$ AU, the PUI distribution shows a rapid drop at around $430$ km s$^{-1}$, the bulk velocity of SW inside the TS, indicating that most particles are injected with the speed of the SW $V_{SW}$ in the plasma frame. Some particles have filled in the shell at low energies due to adiabatic cooling. Since particles with speeds $v > V_{SW}$ result from local acceleration in turbulent flux tubes, the mixture of core PUIs and tail PUIs yields this step like feature \citep{1996GeoRL..23.2871S}. These distributions are to be interpreted in the time-averaged sense, over many flux tubes passing an observer.

A power-law suprathermal tail develops at $v > V_{SW}$. The tail extends to higher energies with increasing radial distance. In view of the flux tube picture, an observer would alternately see distributions with more or less developed tails; Figure \ref{fig:1} is their average over several days or even months (the model is not time dependent, therefore the averaging interval can be arbitrary long). A hump at around $100$ km s$^{-1}$ at $100$ AU is clearly seen, which consists of the low-energy PUIs created in the inner heliosheath. The hump increases in height as the flow slows down toward the heliopause due to the accumulation of low-energy PUIs produced in the heliosheath.

The right panel of Figure \ref{fig:1} illustrates the variation of the PUI VDF along the polar direction ($\theta=0^{\circ}, \phi=0^{\circ}$). The principle features are similar to the equatorial case. At $r = 5$ AU, the PUI distribution shows a sharp change in slope around $725$ km s$^{-1}$, the bulk velocity of the fast SW inside the TS. A pronounced hump develops only beyond the TS (at about $90$ AU). In both panels one can see that the PUI gas gets compressed by the shock and an accelerated power law tail develops out of the PUI core. There is little change in the power-law tail in the heliosheath, which means stochastic acceleration cannot produce a spectrum that is any harder. The downstream value of the power law index is about $-5.2$ in both direction.

\begin{figure}
\epsscale{0.8}
\plotone{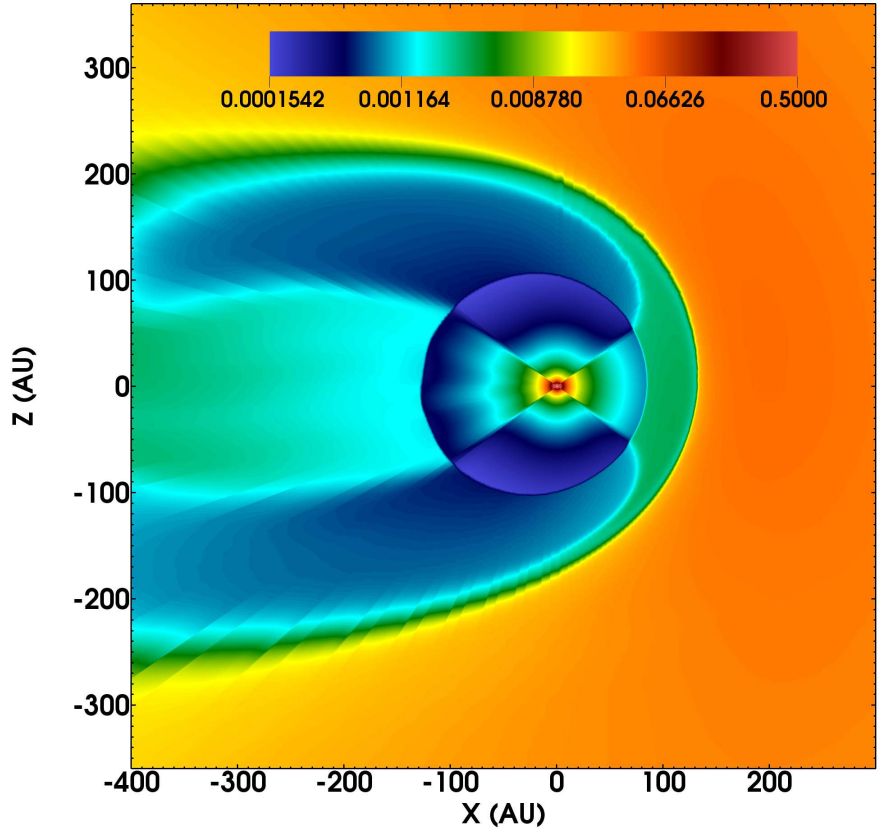}
\caption{Plasma background density (cm$^{-3}$) in the meridional plane. Distinct bands of fast and slow wind in the high and low latitude regions are shown. The slow wind latitudinal extent angle is assumed to be 36$^\circ$. Fast SW prevailing at higher latitudes produces denser subsonic wind in the heliosheath.}
\label{fig:2}
\end{figure}

\begin{figure}
\epsscale{0.5}
\subfloat[]{\includegraphics[width = 2.3in]{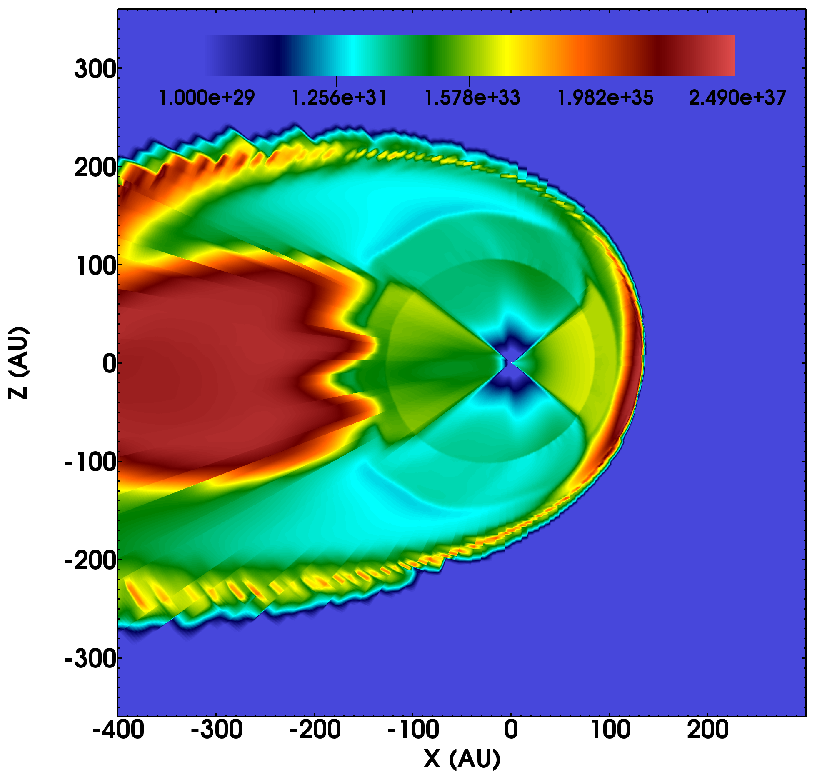}}
\subfloat[]{\includegraphics[width = 2.3in]{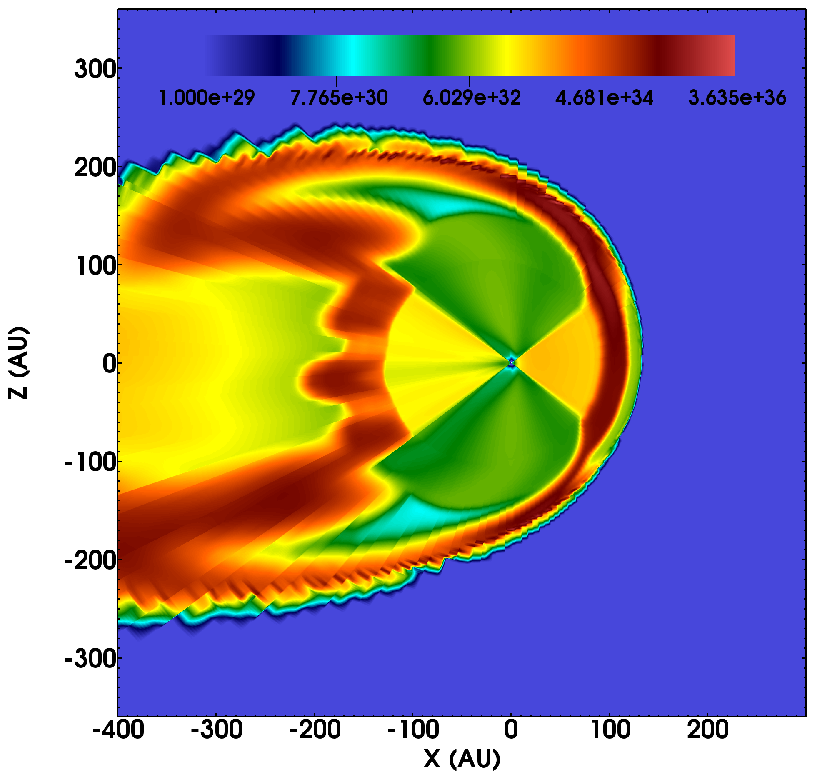}}
\subfloat[]{\includegraphics[width = 2.3in]{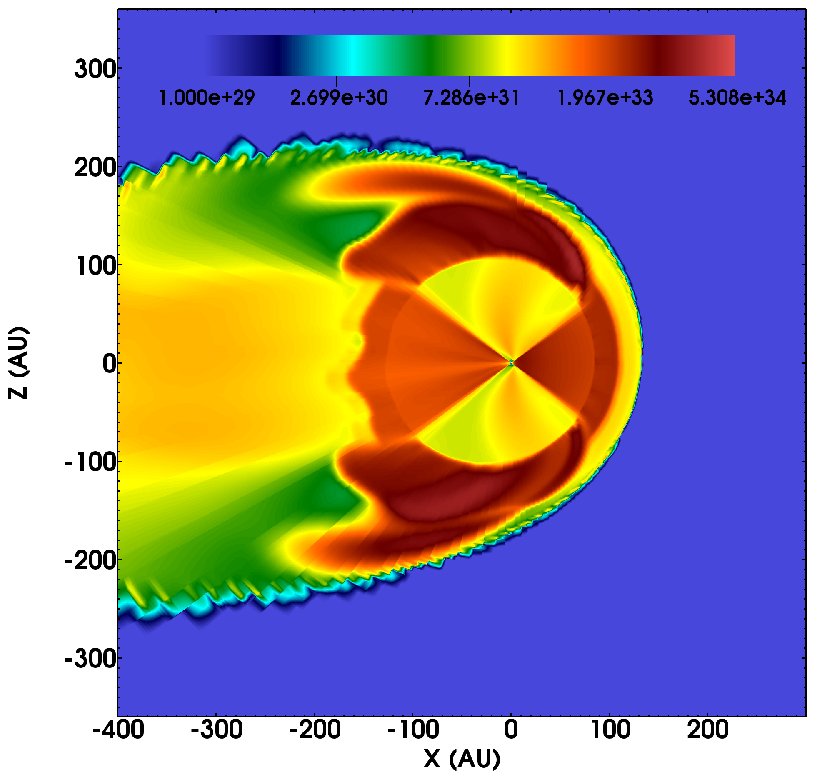}}\
\subfloat[]{\includegraphics[width = 2.3in]{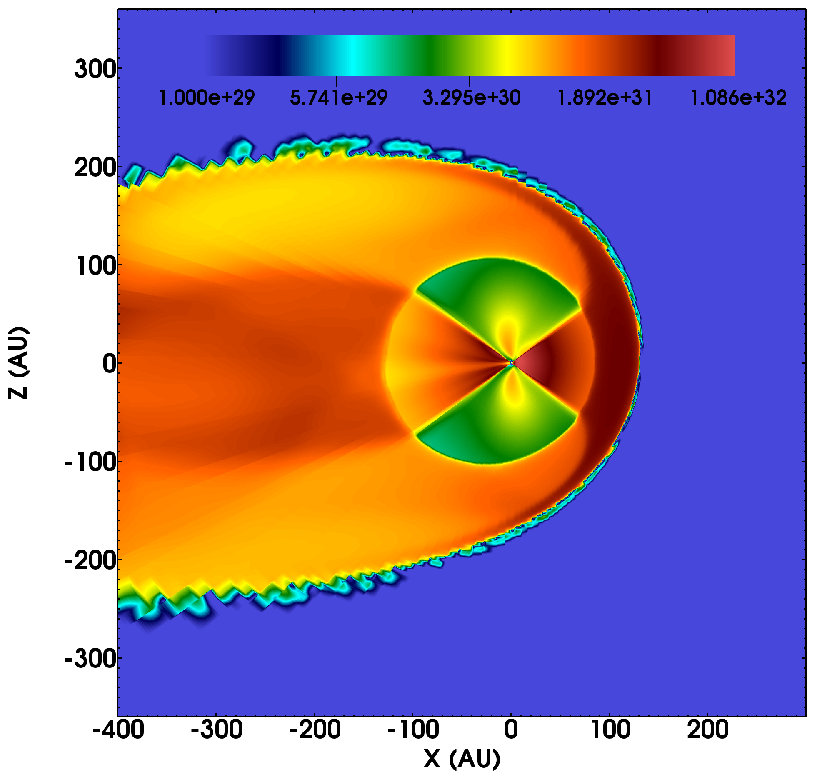}}
\subfloat[]{\includegraphics[width = 2.3in]{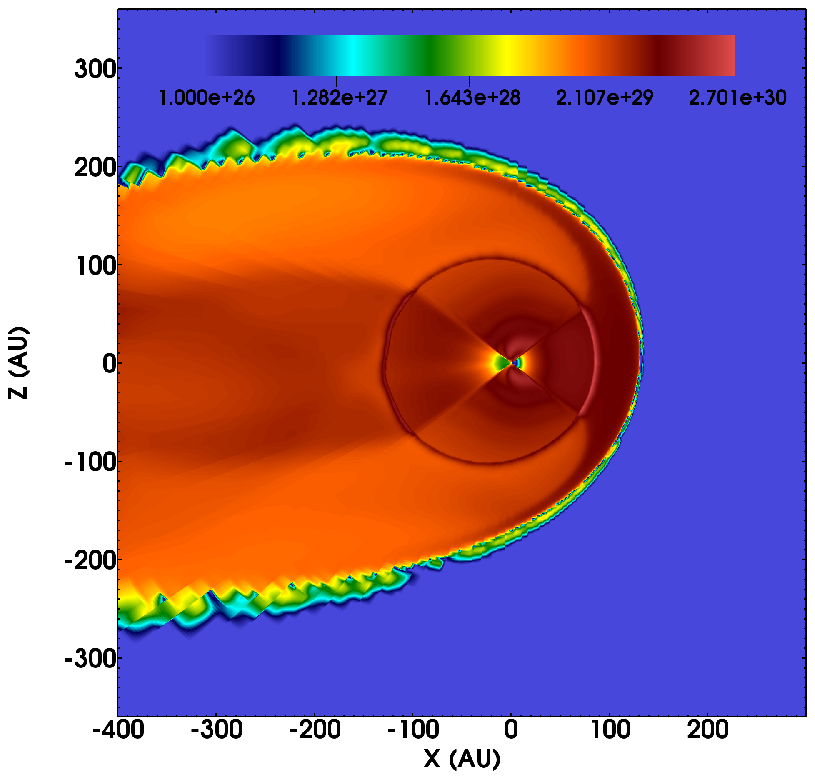}}
\caption{Simulated plasma-frame PUI spatial distributions at (a) $v=30$ km/s, (b) $100$ km/s, (c) $200$ km/s, (d) $v=200$ km/s, and (e) $800$ km/s
in the meridional plane.}
\label{fig:3}
\end{figure}

The distribution of background plasma density is shown in Figure \ref{fig:2}; while Figure \ref{fig:3} presents the core and tail PUI spatial distributions. They show that from the inner boundary to the TS, the PUI density falls off slower than that of the SW core since PUIs are produced over the entire SW. One can clearly see that there is a great difference in PUI densities in the slow SW and in the fast SW. Figures \ref{fig:3}(a) to \ref{fig:3}(c) show that at low energies the PUI density is higher in the slow SW than in the fast SW. The majority of the maps show an increase in PUI densities across the TS due to compression. Naturally, PUIs created in a slow wind have energy lower than those produced in a fast wind. One can see that most low-energy PUIs are in the heliotail from Figure \ref{fig:3}(a). This is because the PUIs produced in the direction of the heliotail have a lower energy on average because of greater SW slowdown by mass loading before the TS.

\begin{figure}
\epsscale{0.8}
\plotone{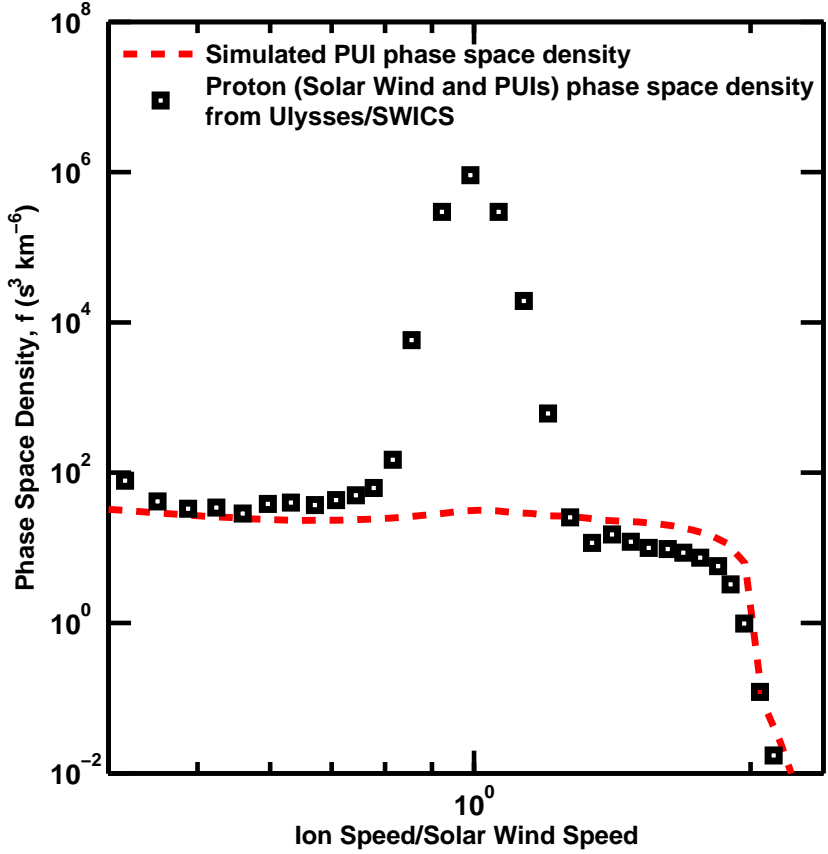}
\caption{Phase space density of H$^{+}$ (including SW and PUIs) versus ion speed in the spacecraft frame. Individual data points are SWICS observation in a 100-day interval in 1994. Red line are the model result at about $3.0$ AU near the southern pole. The SW peak is not shown in the simulation result.}
\label{fig:4}
\end{figure}

The model is tuned to fit the observations. For example, \cite{1998SSRv...86..127G} reported an averaged phase space density measured by SWICS in a 100-day interval in 1994 when {\it Ulysses} was at about 3 AU near the southern pole. To compare the observed velocity distribution with model predictions, we transformed our model distribution functions into the frame of SWICS, and then integrated over the SWICS field of view. The results are presented in Figure \ref{fig:4}. The red line is the simulated PUI phase space density after transformation. One can see that our model result matches the observed phase space density well.

\begin{figure}
\epsscale{0.5}
\subfloat[]{\includegraphics[width = 2.8in]{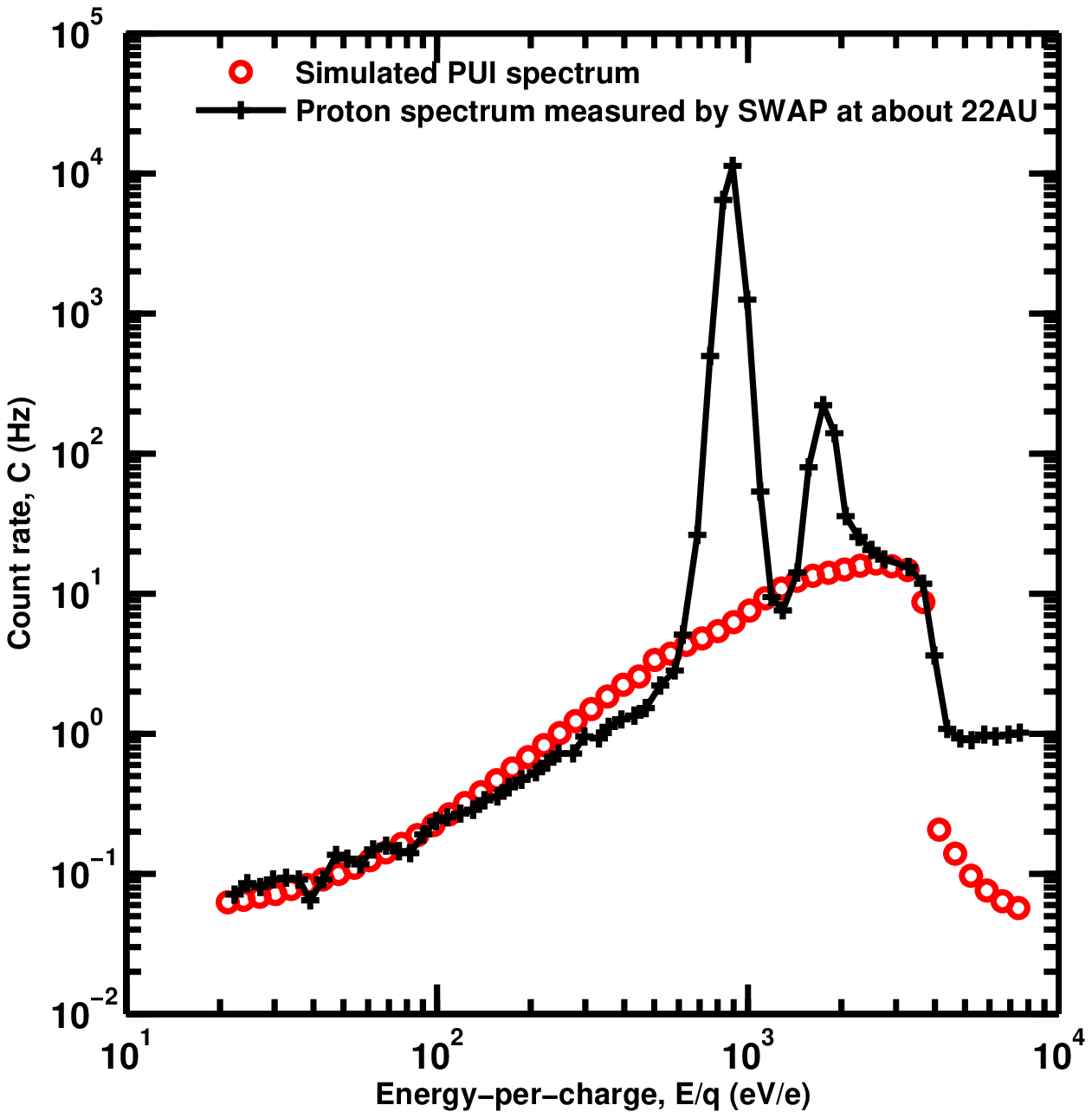}}
\subfloat[]{\includegraphics[width = 2.8in]{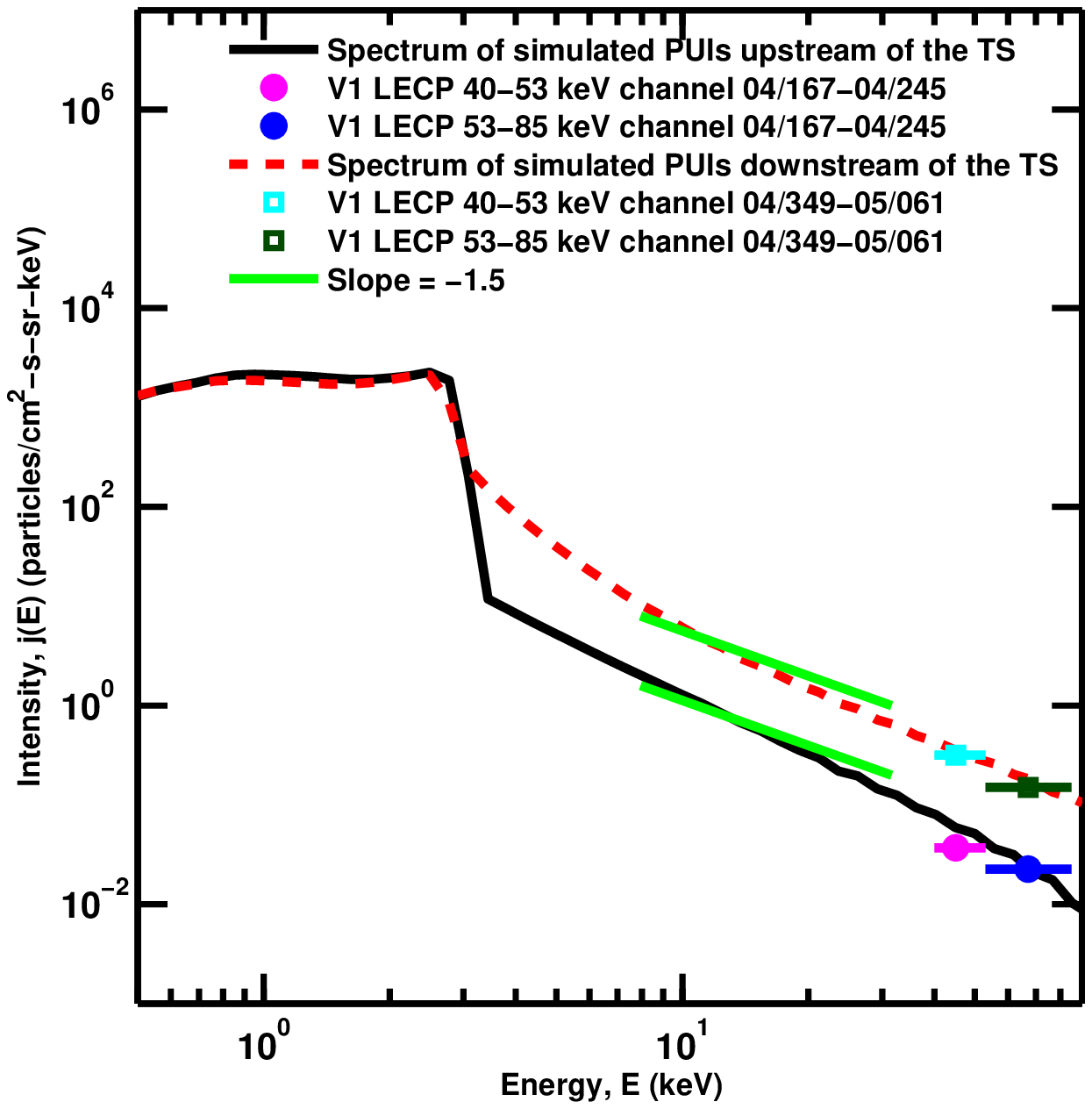}}\
\subfloat[]{\includegraphics[width = 2.8in]{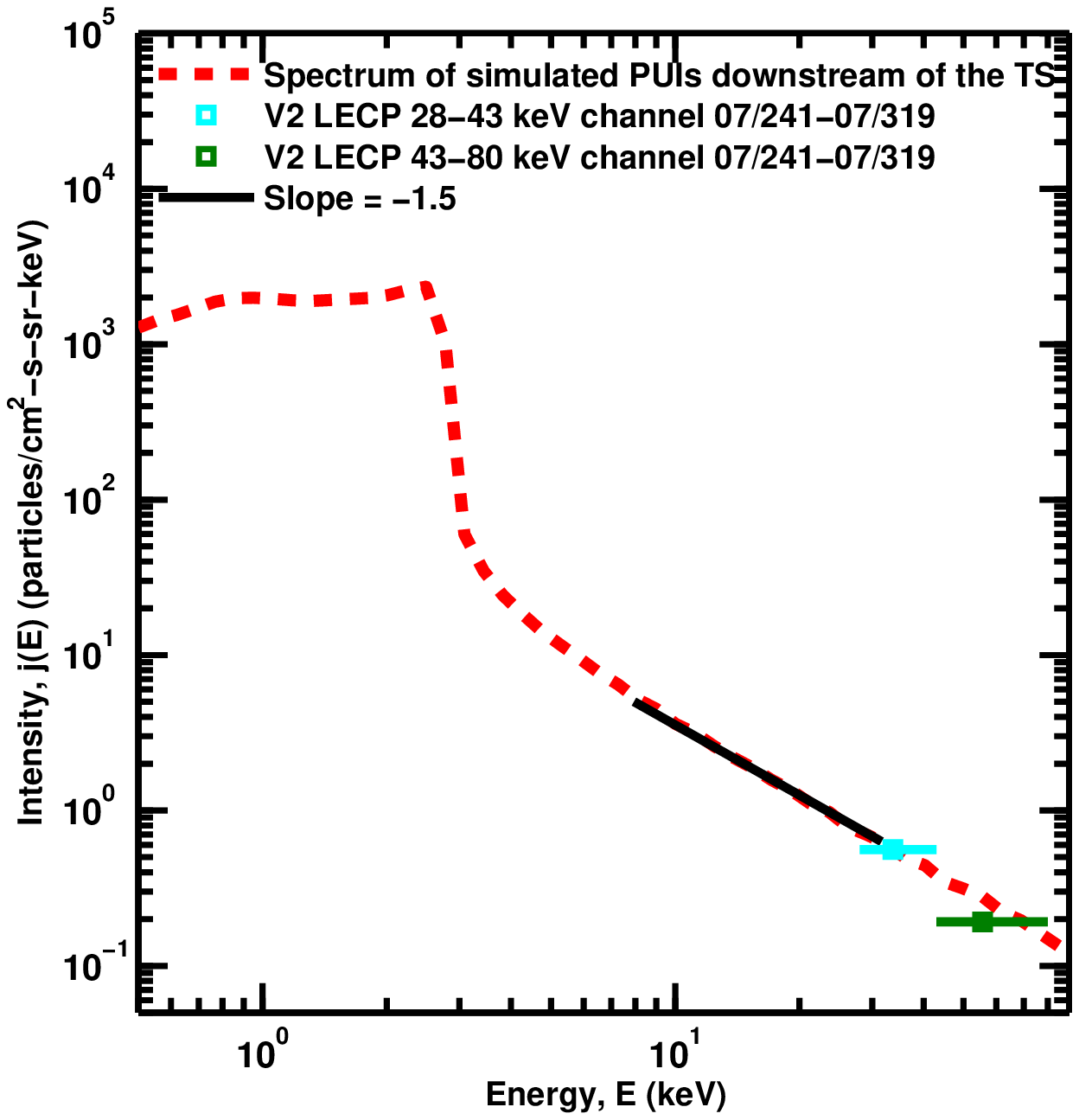}}
\subfloat[]{\includegraphics[width = 2.8in]{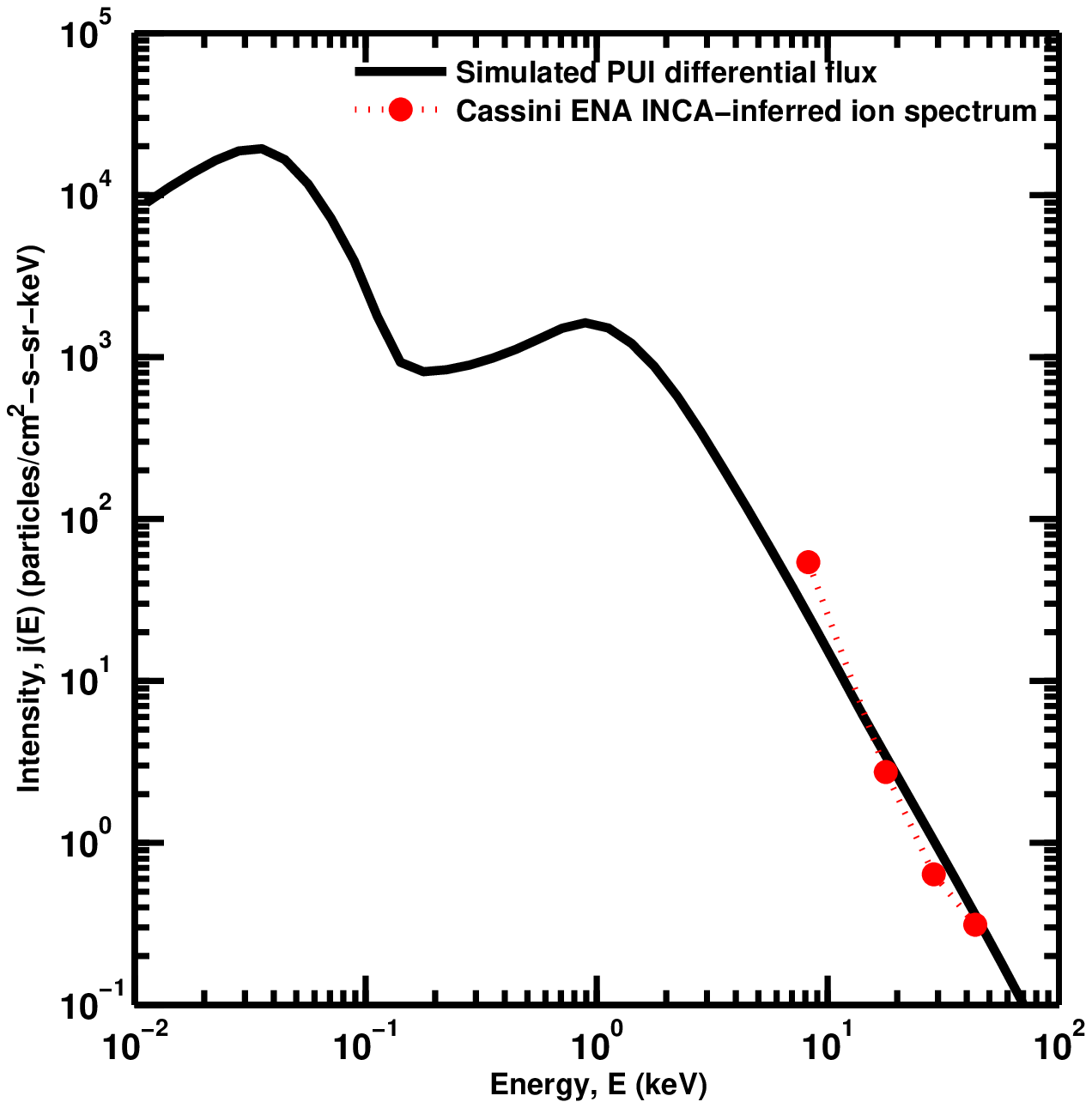}}
\caption{(a) Coincidence count rate of SW and interstellar PUIs versus E/q spectra observed by SWAP at $11.44$ AU. Red circles are the model result. (b) Energy spectra of simulated PUIs and $40$-$85$ keV ions at {\it Voyager 1}, averaged over selected 78-day periods before (pre-TS) and after (post-TS) TS crossings. (c) Energy spectra of simulated PUIs and $28$-$80$ keV ions at {\it Voyager 2}, averaged over selected 78-day periods in the inner heliosheath. (d) The simulated PUI differential flux at $\sim 90$ AU in {\it Voyager 2} direction compared with the {\it Cassini} ENA INCA-inferred ion spectrum \citep{2010AIPC.1302...79K}.}
\label{fig:5}
\end{figure}

Recent measurements by Solar Wind Around Pluto (SWAP) on board {\it New Horizons} were reported in \cite{2013ApJ...768..120R}. SWAP measured SW and PUI spectra from $11$ AU to $22$ AU. We transformed our model distribution function to the spacecraft frame and converted it into count rate. The comparison is given in Figure \ref{fig:5}(a). The SW H$^{+}$ and He$^{2+}$ peaks can be seen at energy per charge (E/q) $\approx 600$ eV/e and $1100$ eV/e, respectively As can be seen, the slope of the spectrum and the PUI step like feature are well reproduced. At all other energies, the PUIs are hidden beneath the SW background. The spectrum is below the data at $v > V_{SW}$. This is partially due to contamination from SW, but also due to local acceleration of the PUIs by variations in magnetic field or the SW background. 

Figure \ref{fig:5}(b) compares the simulated PUI intensities $j(E)$ with {\it Voyager 1} Low Energy Charged Particle (LECP) proton intensities averaged over one selected 78-day interval in the termination foreshock (2004/167-2004/245) and one 78-day period immediately behind the TS (2004/349-2005/061) \citep{2008AIPC.1039..349D}. We also compare simulated PUI intensity with {\it Voyager 2} low-energy proton intensity averaged over a selected 78-day interval in the inner heliosheath (2007/241-2007/319) in Figure \ref{fig:5}(c). Note that the spectral shapes in Figure \ref{fig:5}(b) and (c) are essentially the same, well represented by $j \propto E^{-1.5}$ corresponding to $f \propto v^{-5}$, with the {\it Voyager 2} ion spectrum being slightly harder than that at {\it Voyager 1} behind the TS crossing. Each of the three spectra have the step feature. The spectra shift upward once the TS is crossed.

A partial overlap exists between the {\it Voyager 1, 2} LECP ion instrument ($28 <$ E $< 4000$ keV) and the {\it Cassini} ENA Ion and Neutral Camera (INCA) sensor ($5.2 <$ E $< 55$ keV) \citep{2010AIPC.1302...79K}. \cite{2010AIPC.1302...79K} reported the INCA-inferred ion spectrum in the heliosheath and matched it to the in situ measured {\it Voyager 2} spectrum. Figure \ref{fig:5}(d) shows our simulated PUI spectrum at $\sim 90$ AU in {\it Voyager 2} direction compared with the {\it Cassini} ENA INCA-inferred ion spectrum.

\section{DISCUSSION}
Our model introduced several improvements over the existing models. \cite{2004AdSpR..34...99C} have investigated the spatial variation of PUI spectra, but only the upwind part of the heliosheath was considered. \cite{2006A&A...445..693M} have performed a multi-species simulation, but the magnetic field was ignored and their model was two-dimensional. The model presented here computes PUI distributions on a three dimensional grid. \cite{2006JGRA..111.7101U} considered the SW outside $1$ AU as a combination of three co-moving species, SW protons, electrons, and PUIs, but they only computed the global structure of the SW from the coronal base to $100$ AU without the TS. Our model's external boundary is well in the LISM covering supersonic SW region, the inner heliosheath and the outer heliosheath.

Several weaknesses of the present model are now pointed out. Firstly, the two-population assumption is admittedly questionable. The \cite{1996GeoRL..23.2871S} paper reported a range of field strength variations, rather than a bimodal distribution. The observed distribution functions are averaged over longer times than the averaging interval in \cite{1996GeoRL..23.2871S}, and therefore are a product of a superposition of high and low values of $\eta^2$. The simulated spectra shown in Figures 1 and 5 should be compared with monthly or even yearly averages of the data. Secondly, our model assumes the PUI VDF is isotropic, which is not always the case. For example, ion angular data from {\it Voyager 1} observations during 2002.58 to 2003.10, $85.3$ to $87.3$ AU showed large beamlike anisotropies \citep{2005Sci...309.2020D}. Thirdly, our model is time independent. Incorporating time-dependent SW boundary conditions may improve the results and produce better agreement with the observations. Finally, charge exchange of the PUIs on interstellar H atoms was ignored. This process replaces one pickup ion with another, drawn from a different velocity distribution. While, in principle, including the loss of PUIs would be straightforward, the production term requires numerical integration over the VDF of PUIs in each computational cell, which is a costly procedure. PUI charge exchange may be important in the inner heliosheath, causing an energy redistribution in their VDF.

In spite of these limitations, our model does provide insights into the interpretation of the PUI data and may be used to predict PUI distribution at all locations inside the heliosphere. These distributions show the details that are directly comparable with those seen in spacecraft data. We have obtained the rapid drops in the spectra that appear to be required to match the observations. The model also features power-law tails in the energy, which are commonly observed in space. A velocity diffusion origin of these tails appears to be a valid interpretation.

The compressed SW and PUIs behind the TS create energetic neutral atoms (ENAs) via charge exchange. ENAs with energies high enough to overcome the outward flow speed can be directed back at Earth. Future work will use these PUI results to calculate ENA fluxes at $1$ AU. We plan to compare the simulated ENA fluxes with the {\it IBEX} distributed ENA sky maps. This will bring us closer to explaining why the distributed ENA flux spectrum does not show a knee and why is it close to a power law \citep{2011ApJ...731...56S}.

\acknowledgments
This work was supported by NASA grant NNX12AH44G.

\end{document}